\documentclass[aps, prd, twocolumn, notitlepage, nofootinbib,superscriptaddress]{revtex4-1}

\usepackage{amsfonts,amssymb,amsmath}
\usepackage{gensymb}
\usepackage{color}
\usepackage{graphicx}
\usepackage{hyperref}

\graphicspath{{./}}
\usepackage[normalem]{ulem}

\newcommand{\greaterthanapprox}{\mathrel{\vcenter{
  \offinterlineskip\halign{\hfil$##$\cr
    >\cr\noalign{\kern2pt}\sim\cr\noalign{\kern-2pt}}}}}
    
\newcommand{\nn}{\nonumber}

\newcommand{\WJ}[6]{\left(\begin{array}{c c c}#1&#2&#3\\#4&#5&#6\end{array}\right)}

\DeclareMathOperator{\Tr}{Tr}

\newcommand{\be}{\begin{equation}}        
\newcommand{\ee}{\end{equation}}
\newcommand{\ba}[1]{\begin{align}
#1
\end{align}}

\newcommand{\lb}{\left(}
\newcommand{\rb}{\right)}
    
\newcommand{\ti}{\textit}

\begin{document}

%\title{Something KSZ Something CIB}
\title{Remote Dipole Field Reconstruction with Dusty Galaxies}

\author{Fiona McCarthy}
\email{fmccarthy@perimeterinstitute.ca}

\affiliation{Perimeter Institute for Theoretical Physics, Waterloo,
Ontario, N2L 2Y5, Canada}
\affiliation{Department of Physics and Astronomy, University of Waterloo, Waterloo, Ontario, Canada, N2L 3G1}

\author{Matthew C. Johnson}
\email{mjohnson@perimeterinstitute.ca}
\affiliation{Department of Physics and Astronomy, York University, Toronto, Ontario, M3J 1P3, Canada}
\affiliation{Perimeter Institute for Theoretical Physics, Waterloo,
Ontario, N2L 2Y5, Canada}

\date{\today}

%%%	ABSTRACT	%%%
\begin{abstract}
\noindent
The kinetic Sunyaev Zel'dovich (kSZ) effect, cosmic microwave background (CMB) temperature anisotropies induced by scattering of CMB photons from free electrons, forms the dominant blackbody component of the CMB on small angular scales. Future CMB experiments on the resolution/sensitivity frontier will measure the kSZ effect with high significance. Through the cross-correlation with tracers of structure, it will be possible to reconstruct the remote CMB dipole field (e.g. the CMB dipole observed at different locations in the Universe) using the technique of kSZ tomography. In this paper, we derive a quadratic estimator for the remote dipole field using the Cosmic Infrared Background (CIB) as a tracer of structure. We forecast the ability of current and next-generation measurements of the CMB and CIB to perform kSZ tomography. The total signal-to-noise of the reconstruction is order unity for current datasets, and can improve by a factor of up to $10^3$ for idealised future datasets, based on statistical error only. The CIB-based reconstruction is highly correlated with a galaxy survey-based reconstruction of the remote dipole field, which can be exploited to improve constraints on cosmological parameters and models for the CIB and distribution of baryons.
 \end{abstract}

\maketitle

%%%	INTRODUCTION	%%%
\section{Introduction}
%\begin{itemize}
%\item What is the kSZ effect?
%\item Why is it interesting?
%\item What are WE doing?
%\item WHY??
%\end{itemize}

\noindent
The Cosmic Microwave Background (CMB) temperature anisotropies have been measured with ever-improving angular resolution and sentitivity, from COBE's~ \cite{1992ApJ...396L...1S,1996ApJ...464L...1B} measurements of $\sim 10 \mu K$ fluctuations on angular scales of $\sim 10\degree$, to the Planck satellite \cite{Akrami:2018vks} and ground-based CMB experiments such as SPT and ACT measuring $\sim 10^{-2} \mu K$ fluctuations on $\sim 1'$ scales. These experiments have extracted almost all of the information from the \ti{primary} CMB --- anisotropies sourced primarily at the surface of last scattering, where the CMB was released. However, there remains much information to be extracted from CMB \ti{secondaries}: anisotropies generated through the interaction of CMB photons with mass (lensing) or charges (the Sunyaev Zel'dovich effect) throughout the Universe. These effects are on the resolution/sensitivity frontier, and while they have been detected with moderate significance thus far, future experiments such as Simons Observatory~\cite{Ade:2018sbj}, CCAT-p~\cite{2018SPIE10700E..5XP}, CMB-S4~\cite{Abazajian:2016yjj}, PICO~\cite{Hanany:2019lle}, or CMB-HD~\cite{Sehgal:2019nmk} will provide highly significant measurements of the secondary CMB.

The kinetic Sunyaev Zel'dovich (kSZ) effect is one such secondary temperature anisotropy, induced by the scattering of CMB photons off electrons with non-zero CMB dipole in their rest frame \cite{1980MNRAS.190..413S}. The kSZ effect is the dominant blackbody component of the CMB on angular scales corresponding to $l\agt 4000$, and has been detected at the $>4\sigma$ level~\cite{2012PhRvL.109d1101H,Ade:2012nia,Schaan:2015uaa,Soergel:2016mce,Hill:2016dta,DeBernardis:2016pdv,George:2014oba}. The kSZ effect is interesting from a cosmological perspective because of its dependence on the remote dipole field, the CMB dipole observed at different locations in the Universe. The remote dipole field can be reconstructed using the correlations between CMB temperature and a tracer of density, a technique known as \ti{kSZ tomography} \cite{Zhang:2000wf,2009arXiv0903.2845H,2011MNRAS.413..628S,Munshi:2015anr,Zhang:2015uta,Terrana:2016xvc,Deutsch:2017ybc,Cayuso:2018lhv,Smith:2018bpn}. Forecasts \cite{Deutsch:2017ybc,Cayuso:2018lhv,Smith:2018bpn} indicate that the remote dipole field can be reconstructed with high signal-to-noise through cross-correlation of next-generation CMB experiments and a large redshift survey such as LSST \cite{2009arXiv0912.0201L}. These measurements of the remote dipole field have the potential to improve constraints on primordial non-gaussianity~\cite{Munchmeyer:2018eey}, determine the physical nature of various anomalies in the primary CMB~\cite{Cayuso:2019hen} (e.g. the lack of power on large scales, the hemispherical power asymmetry, the alignment of low multipoles), provide precision tests of gravity~\cite{Contreras:2019bxy,Pan:2019dax}, and constrain the state of the Universe before inflation~\cite{Zhang:2015uta}. 

In this paper we adapt the techniques of Ref.~\cite{Deutsch:2017ybc} to define a quadratic estimator for the remote dipole field from a \ti{2-dimensional} tracer of structure in cross-correlation with CMB temperature maps. We apply this estimator to the Cosmic Infrared Background (CIB), infrared radiation from dusty star-forming galaxies. The anisotropies in the CIB trace the distribution of large scale structure (LSS), and have been measured with increasing accuracy from balloon (e.g. BLAST~\cite{2009ApJ...707.1766V}) and ground-based facilities (e.g. SPT~\cite{2010ApJ...718..632H,George:2014oba} and ACT~\cite{Dunkley:2013vu}) as well as satellite missions (e.g. Herschel~\cite{2013ApJ...772...77V} and Planck~\cite{2014A&A...571A..30P}) over a wide range of frequencies ($\sim 100-1200 \ {\rm GHz}$). In principle, the CIB can provide constraints on cosmology. However, the main obstruction to using the CIB as a competitive cosmological probe (e.g. to measure primordial non-Gaussianity~\cite{Tucci:2016hng}) is the lack of maps on large fractions of the sky with sufficiently low foreground residuals (see e.g.~\cite{Lenz:2019ugy}). Although, by virtue of its large correlation with the lensing potential~\cite{Holder:2013hqu}, the CIB has proven to be a useful tool for de-lensing CMB polarization maps to obtain improved constraints on primordial gravitation waves~\cite{Manzotti:2017net}. 

To obtain a high fidelity reconstruction of the remote dipole field, it is necessary to measure the clustered component of the tracer (here the CIB) on angular scales of $\ell \sim 3000-4000$, where kSZ becomes comparable in amplitude to the primary CMB~\cite{Deutsch:2017ybc,Smith:2018bpn}. This resolution/sensitivity has been achieved for the CIB with existing experiments, e.g. SPT~\cite{George:2014oba}, albeit on small areas of the sky. The CIB at a fixed frequency samples structure over a wide range of redshifts, and only a significantly coarse-grained reconstruction of the remote dipole field can be obtained using kSZ tomography. This coarse-grained dipole field has structure primarily on large angular scales, which means that measurements of the CIB on large fractions of the sky are necessary for kSZ tomography. If future measurements can achieve this, the course-grained remote dipole field provided by kSZ tomography will probe the homogeneity of the Universe on the very largest scales. This can be used to constrain various models of early-Universe physics~\cite{Cayuso:2019hen,Zhang:2015uta}.

To assess the utility of the CIB for kSZ tomography, we perform a set of simplified forecasts for the Planck satellite and for a future experiment with specifications similar to a stage-4 CMB experiment~\cite{Abazajian:2016yjj}. In our forecasts, we assume that the CIB can be perfectly separated from the blackbody component (the lensed primary CMB and kSZ) in each channel, as well as from other foregrounds. We further assume data on the full sky, gaussian beam, and white instrumental noise. We perform a principal component analysis to identify which modes of the CMB dipole field we can hope to reconstruct from the CIB and find that reconstruction of a linear combination (in redshift space) of the remote dipole will be possible for modes with redshift-weightings corresponding to that of the CIB integration kernel. Under these assumptions, we find that the remote dipole can be reconstructed with an overall signal-to-noise of order one using Planck-quality data. We also perform a forecast for a next-generation experiment; for an idealistic case with no foregrounds, we find that a full-sky survey could in principle perform a mode-by-mode reconstruction of the remote dipole field on large-scales with high signal-to-noise ($>\mathcal{O}(10)$ for $\ell < 10$), and obtain an overall signal-to-noise of $\mathcal{O}(1000-10000)$. This strongly motivates a detailed study for specific instruments, including realistic foregrounds and systematics, which we defer to future work. 

The remote dipole field can also be reconstructed using future galaxy redshift surveys. We demonstrate that the CIB-based reconstruction of the remote dipole field is highly correlated with the galaxy-based reconstruction. While this implies that there is limited additional cosmological information to mine from the CIB-based reconstruction, the methods developed in this paper CIB information are interesting nevertheless. Firstly, as we wait for LSST-like galaxy surveys to become available to use for tomography, alternative methods are worth investigating as the density tracer maps might become available earlier than LSST. Secondly, the large correlation can be used to address the optical depth degeneracy~\cite{Battaglia:2016xbi,Madhavacheril:2019buy,Smith:2018bpn}: the modelling uncertainty in the correlation between electrons and the tracer used for the reconstruction. The optical depth degeneracy manifests as a redshift-dependent bias on the reconstructed dipole field, different for each tracer. By correlating the CIB-based and galaxy-based reconstructions, it is possible to measure the ratio of optical depth bias parameters with arbitrary accuracy. Such a measurement could both help extract cosmology from reconstructions of the remote dipole field, as well as provide insight into physical models of both the CIB and electron distribution.

The paper is organized as follows. In Section \ref{sec:reconstruction_estimators} we derive the quadratic estimators we use; in Section \ref{sec:forecasts} we will describe our forecast; in Section \ref{sec:forecast_results} we will present the results of our forecasts for Planck-quality data and a next-generation experiment; in Sec.~\ref{sec:correlations} we explore the correlations between the CIB-based and galaxy-based remote dipole reconstructions; we conclude in section \ref{sec:conclusions}. The code we used for our computations is available at \url{https://github.com/fmccarthy/ksz_CIB_forecasts}.

%%%	SECTION 1	%%%
\section{kSZ Tomography: Reconstruction Via a 2-Dimensional Field}\label{sec:reconstruction_estimators}

\noindent
We wish to reconstruct the remote dipole field by cross-correlating the CMB temperature anisotropies with the CIB intensity, which is a two-dimensional field defined by the line-of-site integral over the 3-dimensional  CIB emissivity density. Ref. \cite{Deutsch:2017ybc} derived a minimum variance quadratic estimator for the remote dipole field using a 3-dimensional tracer (galaxy redshifts) and the CMB temperature anisotropies. Here, we adapt this analysis for the 2-dimensional case. Following \cite{Deutsch:2017ybc}, we use the cross-correlation between the kSZ-induced CMB temperature and our tracer to derive this estimator. 

The kSZ-induced temperature anisotropy in the direction $\hat{\mathbf{  n}}$ is given by
\be
\left. \frac { \Delta T } { T } \right| _ { \mathrm { kSZ } } \left( \hat { \mathbf { n } }  \right) =  \int_{0}^{\chi_{re}} d \chi  \ \dot{\tau} \left( \hat { \mathbf { n } } , \chi \right) v _ { \mathrm { eff } } \left( \hat { \mathbf { n } } , \chi \right)
\ee
The integral over comoving radial distance $\chi$ is done out to reionization at $\chi_{re}$. The remote dipole field $v _ { \mathrm { eff } } \left( \hat { \mathbf { n } } , \chi \right)$ is defined by
\be
v _ { \mathrm { eff } } \left( \hat { \mathbf { n } }  , \chi  \right) \equiv \sum_{m=-1}^{1} \Theta_{1}^m \left( \hat { \mathbf { n } } , \chi \right) Y_{1m} \left(\hat { \mathbf { n } } \right),
\ee
\be 
\Theta_{1}^m \left( \hat { \mathbf { n } } , \chi  \right) \equiv \int \ d^2 \hat{ \mathbf { n } }'  \ \Theta(\hat { \mathbf { n } }, \chi, \hat { \mathbf { n } }') Y_{1m}(\hat { \mathbf { n } }')
\ee
where $\Theta(\hat { \mathbf { n } }, \chi,\hat { \mathbf { n } }')$ is the CMB temperature the electron at $(\hat { \mathbf { n } }, \chi)$ sees along direction $\hat { \mathbf { n } }'$, $\Theta(\hat { \mathbf { n } }, \chi,\hat { \mathbf { n } }') = \Theta_{\rm SW} + \Theta_{\rm ISW} + \Theta_{\rm Dop} $, with the usual contributions from Sachs-Wolfe (SW), Integrated Sachs-Wolfe (ISW), and Doppler (Dop) effects. The dominant contribution to the remote dipole field at any point in spacetime is the Doppler effect from the peculiar velocity field, and therefore one can approximate $v _ { \mathrm { eff } } \simeq \vec{v} \cdot \hat { \mathbf { n } }$ unless long-distance correlations of the remote dipole field are considered. Such correlations will be relevant to our discussion below, and so we consider all contributions to the remote dipole field. A more detailed description of the properties of the remote dipole field can be found in Refs.~\cite{Terrana:2016xvc,Deutsch:2017ybc}. The differential optical depth $\dot\tau(\chi,\hat{\mathbf{ n}})$ is defined by
\be
\dot\tau\lb\chi, \hat{\mathbf{ n}}\rb=-\sigma_ T a(\chi) n_e\lb\chi, \hat{\mathbf{ n}}\rb;\label{taudot}
\ee
here $\sigma_T$ is the Thompson scattering cross section, which governs the rate of the scattering of the photons and electrons; $a(\chi)$ is the scale factor at comoving distance $\chi$; $n_e\lb\chi, \hat{\mathbf{ n}}\rb$ is the comoving electron number density.

We proceed by defining $N$ redshift bins, labelled by $\alpha$ where $\alpha\in 1\cdots N$, each with comoving-distance boundaries $\chi^\alpha_{min},\chi^\alpha_{max}$. The optical depth $\tau^\alpha$ in each bin is thus defined as
\be
\tau^\alpha(\hat{\mathbf{ n}}) \equiv \int _{\chi^\alpha_{min}}^{\chi^\alpha_{max}}d\chi\dot\tau\lb\chi, \hat{\mathbf{ n}}\rb.
\ee
The bin-averaged dipole field is
\be
\bar v_{eff}^\alpha(\hat{\mathbf{ n}})=\frac{1}{\chi^\alpha_{max}-\chi^\alpha_{min}}\int _{\chi^\alpha_{min}}^{\chi^\alpha_{max}} d\chi\, v_{eff}\lb\chi, \hat{\mathbf{ n}}\rb
\ee
with
\be
 v_{eff}\lb\chi^\alpha, \hat{\mathbf{ n}}\rb=\bar v_{eff}^\alpha(\hat{\mathbf{ n}})\lb1+\delta  v_{eff}\lb\chi^\alpha, \hat{\mathbf{ n}}\rb\rb.
\ee
Contributions from $\delta v_{eff}$ are small, due to cancelations along the line-of-sight, and we therefore have 
 \be
\frac{\Delta T}{T}(\hat{\mathbf{ n}}) \simeq \sum_\alpha \tau^\alpha(\hat{\mathbf{ n}}) \bar {v}_{eff}^\alpha(\hat{\mathbf{ n}}) .\label{t_tau_v}
\ee

The CIB brightness $I(\nu,\hat{\mathbf{ n}})$ at frequency $\nu$ is given by a line-of-sight integral over emissivity density $j_\nu(\chi, \hat{\mathbf{ n}})$ (see e.g.~\cite{2001ApJ...550....7K}):
\be\label{CIBbrightness}
I (\nu, \hat{\mathbf{ n}})= \int _{0}^{\chi_{re}} d\chi\, a(\chi )j_\nu\lb\chi, \hat{\mathbf{ n}}\rb.
\ee
To model the observed CIB brightness, we use the halo model of \cite{2012MNRAS.421.2832S} and follow the ``minimally empirical'' model for the mean emissivity density of \cite{Wu:2016vpb}; for further details see Appendix \ref{App:halo_CIB}. 

We now compute the cross-correlation bewteen the kSZ temperature anisotropy \eqref{t_tau_v} and CIB brightness \eqref{CIBbrightness}. We work in spherical harmonic space with the conventions
\ba{
O(\hat{\mathbf{ n}}) &=\sum_{lm}  o_{lm}Y_{lm}( \hat{\mathbf{ n}});\\
 o_{lm}&= \int d^2  \hat{\mathbf{ n}} \, O(\hat{\mathbf{ n}}) Y^*_{lm}( \hat{\mathbf{ n}});
}
$o_{lm}$ are the coefficients of the expansion, which we denote for temperature as $a_{lm}^T$ and CIB intensity as $I_{lm}$. The cross-correlation is given by
\ba{
\left< a_{l_1m_1}^T I_{l_2m_2}\right>=&\sum_{\alpha}\sum_{L_1M_1L_2M_2}\bar v_{L_2M_2}^\alpha \left<\tau^\alpha_{L_1M_1}I_{l_2m_2}\right>\nn\\
&\int  d^2  \hat{\mathbf{ n}} \, Y^*_{l_1m_1}( \hat{\mathbf{ n}}) Y_{L_1M_1}( \hat{\mathbf{ n}})Y_{L_2M_2}( \hat{\mathbf{ n}}) .\label{tointegrateovern}
}
We take the $v_{LM}$ term out of the ensemble average because the dominant contribution to $\left<\tau^\alpha_{L_1M_1} \bar v_{L_2M_2}^\alpha I_{l_2m_2}\right>$ comes from $\bar v_{L_2M_2}\left<\tau^\alpha_{L_1M_1} I_{l_2m_2}\right>$~\cite{Terrana:2016xvc}. 
Using statistical isotropy to write angular power $\left<\tau^\alpha_{L_1M_1}I^{*}_{l_2m_2}\right>$ as
\be
\left<\tau^\alpha_{l_1m_1}I^{*}_{l_2m_2}\right>\equiv C_{l_1}^{\tau I\alpha} \delta_{l_1l_2}\delta_{m_1m_2},\label{cltauj}
\ee
or $\left<\tau^\alpha_{l_1m_1}I_{l_2m_2}\right>=(-1)^{m_2} C_{l_1}^{\tau I\alpha} \delta_{l_1l_2}\delta_{m_1,-m_2}$, and performing the angular integration in \eqref{tointegrateovern}, gives us
\begin{widetext}
\ba{
\left< a_{l_1m_1}^TI_{l_2m_2}\right>=&\sum_{\alpha;L_2M_2}(-1)^{m_1+m_2} \sqrt{\frac{(2l_1+1)(2l_2+1)(2L_2+1)}{4\pi}}\lb
\begin{array}{c c c}
l_1&l_2&L_2\\
0&0&0\end{array}
\rb\lb\begin{array}{c c c}
l_1&l_2&L_2\\
-m_1&-m_2&M_2\end{array}
\rb
\bar v_{L_2M_2}^\alpha  C_{l_2}^{\tau I\alpha}
}
\end{widetext}
where $\WJ{l}{l_1}{l_2}{m}{m_1}{m_2}$ are the Wigner 3J symbols. We then define
\be
\Gamma_{l_1l_2l_3}^\alpha\equiv\sqrt{\frac{\lb2l_1+1\rb\lb2l_2+1\rb\lb2l_3+1\rb}{4\pi}}\WJ{l_1}{l_2}{l_3}{0}{0}{0}  C_{l_2}^{\tau I\alpha}\label{Gamma}
\ee
such that
\ba{
\left< a_{l_1m_1}^TI_{l_2m_2}\right>=\sum_{\alpha;L_2M_2}&(-1)^{m_1+m_2}\Gamma_{l_1l_2L_2}^\alpha\nn\\
&\times \lb\begin{array}{c c c}
l_1&l_2&L_2\\
-m_1&-m_2&M_2\end{array}
\rb \bar v_{L_2M_2}^\alpha.\label{cross_correlation}
}
\subsection{Defining an estimator for the dipole field}
\noindent
We want to define a minimum variance quadratic estimator for $\bar v$ as in \cite{Deutsch:2017ybc}. Note, however, the key difference that in \cite{Deutsch:2017ybc} the CMB temperature was being cross-correlated with a 3-D field that \ti{could itself be binned into redshift bins} just as we bin the optical depth, and constrast this to the case we have here, where we cross-correlate the CMB temperature field with a 2-D field that cannot be redshift-binned. This difference manifests itself in the sum over $\alpha$ on the right hand side of \eqref{cross_correlation}.

We can proceed by defining an estimator
\be
\hat v^\alpha_{lm}=\sum_{l_1m_1l_2m_2} W^\alpha_{lml_1m_1l_2m_2}a_{l_1m_1}^TI^{*}_{l_2m_2}
\ee
that minimizes the variance, but dropping the constraint that it must be unbiased; i.e. we have
\be
\left<\hat v^\alpha_{lm}\right>\ne\bar v^\alpha_{lm}.
\ee
We can then build unbiased estimators from linear combinations of $\hat v^\alpha$, using the techniques of \cite{2013MNRAS.431..609N}; we will find that
\be
\left<\hat v^\alpha_{lm}\right>=R_{\alpha\beta}\bar v^\beta_{lm}\label{rotmar}
\ee
for some rotation matrix $R_{\alpha\beta}$. This will allow us to define an \ti{unbiased} quadratic estimator $\hat {v}_{lm}^{\prime\alpha}$ by 
\be
\hat { v}_{lm}^{\prime\alpha}=\lb R^{-1}\rb_{\alpha\beta} \hat v^\beta_{lm}\label{unbiased_estim}
\ee
which will have 
\be
\left<\hat { v}_{lm}^{\prime\alpha}\right>=\bar v^\alpha_{lm}.
\ee
Of course, this will not necessarily be a minimum variance estimator; it will have variance 
\ba{
\left<\hat{ v}_{lm}^{\prime\alpha} \hat { v}_{l'm'}^{\prime\beta} \right>=&\lb R^{-1}\rb_{\alpha\gamma}\lb R^{-1}\rb_{\beta\delta} N^{\bar v \bar v}_{\gamma\delta l} \delta_{ll'}\delta_{mm'}\label{rot_noise}
}
where the variance of the original estimator $\hat v^\alpha_{lm}$ is given by
\be
\left<\hat v^\alpha_{lm}\hat v^\beta_{l'm'}\right>=N^{\bar v \bar v}_{\alpha\beta l}\delta_{ll'}\delta_{mm'}.
\ee

To find the (biased) minimum variance estimator for each bin, we can follow \cite{Okamoto:2003zw} and rewrite the $m$-dependence of the weights $W^\alpha_{lml_1m_1l_2m_2}$ in terms of the Wigner 3J-symbols along with a normalisation $A_l^\alpha$ and $l$-coupling term $w^\alpha _{l_1l_2l}$ such that
\be
\hat v_{lm}^\alpha=A_l^\alpha(-1)^m\sum_{l_1m_1l_2m_2}\WJ{l_1}{l_2}{l}{m_1}{m_2}{-m}w^\alpha _{l_1l_2l} \,a_{l_1m_1}^TI_{l_2m_2};\label{newestim}
\ee
we find (using \eqref{cross_correlation}) that the mean is 
\ba{
\left<\hat v_{lm}^\alpha\right>=\frac{A_l^\alpha}{2l+1}\sum_{\beta;l_1l_2}&w^\alpha _{l_1l_2l} \,\Gamma_{l_1l_2l}^\beta \ \bar v_{lM_2}^\beta.
}
As we have no requirements on what the mean should be, we can choose to ``ignore'' the contributions to the mean from the terms in the sum over $\beta$ where $\beta\ne\alpha$ and, so that we can follow closely the standard procedure of \cite{Deutsch:2017ybc}, define the normalisation $A^\alpha$ as
\be
A_l^\alpha = \frac{2l+1}{\sum_{l_1l_2}w^\alpha_{l_1l_2l}\Gamma_{l_1l_2l}^\alpha }.\label{normalisation_A}
\ee
We can now proceed to minimize the variance of the estimator to solve for $w^\alpha_{l_1l_2l}$; we find that
\be
w_{l_1l_2l}^\alpha=\frac{\Gamma^{\alpha}_{l_1l_2l} }{C_{l_1}^{TT}C_{l_2}^{II}}\label{w_sol},
\ee
where $C_{l}^{TT}$ is the temperature angular power spectrum and $C_{l}^{II}$ is the angular power spectrum of the CIB brightness. Thus the (biased) minimum variance estimator in each bin is
\be
\hat v^\alpha_{lm}=A^\alpha_l(-1)^m\sum_{l_1m_1l_2m_2}\WJ{l_1}{l_2}{l}{m_1}{m_2}{-m}\Gamma^{\alpha}_{l_1l_2l}\frac{a_{l_1m_1}^Tj_{l_2m_2}}{C_{l_1}^{TT}C_{l_2}^{II}}\label{minvar} %define C^II
\ee
where
\be
\frac{1}{A^\alpha_l}=\frac{1}{2l+1}\sum_{l_1l_2}\frac{\Gamma^{\alpha}_{l_1l_2l}\Gamma^\alpha_{l_1l_2l} }{C_{l_1}^{TT}C_{l_2}^{II}}.
\ee
We define the noise $N^{\bar v \bar v}_{\alpha\beta l}$ as the variance in the absence of signal $\left<\hat v_{lm}^\alpha\hat v_{l'm'}^{*\beta}\right>$; we find
\be
N^{\bar v \bar v}_{\alpha\beta l}=\frac{A^\alpha_lA^{\beta}_{l}}{2l+1}\sum_{l_1l_2}\frac{\Gamma^{\alpha}_{l_1l_2l}\Gamma^{\beta}_{l_1l_2l}}{C_{l_1}^{TT}C_{l_2}^{II}}.\label{min_var_noise}
\ee

\subsection{Debiasing $\hat v^\alpha$}

\noindent
$\hat v^\alpha$ is biased since $\left<\hat v^\alpha_{lm}\right>\ne\bar v^\alpha_{lm}$; indeed, we find that its mean is
\be
\left<\hat v^\alpha_{lm}\right>=\dfrac{1}{\sum_{l_1l_2}\lb\dfrac{\Gamma^{\alpha}_{l_1l_2l}\,^2}{C_{l_1}^{TT}C_{l_2}^{II}}\rb
}\sum_{l_1l_2}\frac{\Gamma^{\alpha}_{l_1l_2l}\Gamma_{l_1l_2l}^{\beta}}{C_{l_1}^{TT}C_{l_2}^{II}}\bar v_{lm}^\beta.
\ee
This allows us to read off the ``rotation matrix'' $R_{\alpha\beta}$ of \eqref{rotmar} as
\be
R_{\alpha\beta}=\dfrac{1}{\sum_{l_1l_2}\lb\dfrac{\Gamma^{\alpha}_{l_1l_2l}\,^2}{C_{l_1}^{TT}C_{l_2}^{II}}\rb
}\sum_{l_1l_2}\frac{\Gamma^{\alpha}_{l_1l_2l}\Gamma_{l_1l_2l}^{\beta}}{C_{l_1}^{TT}C_{l_2}^{II}};\label{ralphabeta}
\ee
we can then define an estimator $\hat{ v}^{\prime\alpha}$ via \eqref{unbiased_estim}, with variance given by \eqref{rot_noise}. 

Henceforward we will drop the prime and refer to the set of unbiased estimators as $\hat v^\alpha$.

\section{Signal-to-noise Forecasts}\label{sec:forecasts}

\subsection{Signal Model}\label{sec:signalmodel}
\noindent
Our signal model follows \cite{Terrana:2016xvc}. We wish to compute the bin-averaged dipole field power spectrum $C_l^{\bar v \bar v}$, which is given by
\be
C^{\bar v \bar v}_{\alpha\beta}{}_l=\int \frac{dk}{\lb2\pi\rb^3}k^2 P(k) \bar \Delta _{\alpha l}(k) \bar \Delta_{\beta l}(k)
\ee
with $\bar \Delta _{\alpha l}^v$ the bin-averaged dipole field transfer function
\be
\bar \Delta _{\alpha l}^v(k)= \frac{1}{\chi^\alpha_{max}-\chi^\alpha_{min}}\int _{\chi^\alpha_{min}}^{\chi^\alpha_{max}} d \chi \Delta ^v_l(k,\chi)
\ee
and $\Delta ^v_l(k,\chi)$ defined by
\be
v^\alpha_{lm}=\int \frac{d^3\mathbf{ k}}{\lb2\pi\rb^3}\Delta_l^v(k,\chi)\Psi_i(\mathbf k)Y_{lm}(\hat{\mathbf{ k)}}.
\ee
$\Delta_l^v(k,\chi)$ has contributions from a Sachs--Wolfe (SW), integrated Sachs--Wolfe (ISW), and Doppler term as given in References \cite{Terrana:2016xvc,Deutsch:2017ybc}. As we will see below, it is necessary to include all contributions to properly model the bin-averaged remote dipole reconstructed using the CIB.

\subsection{Noise Model}

\noindent
The reconstruction noise depends on models for $C_l^{TT}$, $C_l^{II}$, and $C_l^{\tau I}$ (which appears in \eqref{min_var_noise} and \eqref{ralphabeta} via $\Gamma^\alpha_{l_1l_2l}$). We assume that the blackbody contribution to the CMB can be perfectly separated from the CIB, and that all foregrounds for both the CMB and CIB can be perfectly removed.

The CMB temperature anisotropy power spectrum $C_l^{TT}$ is a sum of the lensed primary CMB, the kSZ contribution, and the instrumental noise:
\be
C_l^{TT}=C_l^{lensed}+C_l^{kSZ}+N_l^{TT}.
\ee 
$C_l^{lensed}$ is computed with CAMB \cite{Lewis:1999bs} and $C_l^{kSZ}$ is computed using the model described in Ref.~\cite{Smith:2018bpn}. The instrumental noise $N_l$ is given by
\be
N_l=N_T \exp\lb\frac{l(l+1)\Theta^2}{8\ln 2}\rb\label{Instrumental_Noise}
\ee
where $N_T$ is the noise per pixel squared and $\Theta$ is the (Gaussian) beam Full Width at Half Maximum. 

Our model for $C_l^{II}$, which closely follows that of Ref.~\cite{2012MNRAS.421.2832S}, is presented in Appendix \ref{App:halo_CIB}. Within the halo model-based approach we follow (see e.g. Ref.~\cite{Cooray:2002dia} for a review of the halo model), $C_l^{II}$ comprises four terms: a one-halo term, a two-halo term, a ``shot-noise'' term corresponding to self-pairs of galaxies, and an instrumental noise term:
\be
C_l^{II}=C_l^{1 \, halo}+C_l^{2\, halo}+C_l^{Shot\, Noise}+N_l^{II}.
\ee
The instrumental noise is as in \eqref{Instrumental_Noise}. The shot noise depends on the flux cut for removing point sources, which is a function of the resolution and sensitivity of an experiment. The optical depth-CIB correlation function $C_l^{\tau I}$ is also computed within the halo model; details are in Appendix \ref{app:CIB_EL}.

\subsection{Signal-to-noise}\label{sec:s2n_pca}

\noindent
Because the reconstruction noise $N^{\bar v \bar v}_{\alpha\beta l}$ and signal $C^{\bar v \bar v}_{\alpha\beta l}$ are both correlated between redshift bins, we perform a Principal Component Analysis (PCA) to isolate uncorrelated modes $\hat v'^{\alpha}$. The PCA consists of changing our basis from $\hat v^\alpha$ to $\hat v^{\prime\alpha}$ via a linear transformation $A^{\alpha\beta}$ such that ${\bf C}_{ l}^{\prime\bar v^\prime \bar v^\prime}={\bf A} {\bf C}_l^{\bar v \bar v} {\bf A}^{-1}$ is diagonal with entries equal to the signal-to-noise of each principle component. We then define the signal-to-noise per mode of the remote dipole field for each principle component $\alpha$ as
\be
\lb S /N\rb^\alpha_{lm}=\lb\frac{ f_{sky}}{2} \lb C'^{\bar v' \bar v'}_{{\alpha\alpha l}} \rb^2\rb^{\frac{1}{2}};
\ee
The total signal-to-noise for the reconstructed remote dipole field is:
\be
S /N=\lb\sum_{\alpha;l}\frac{ f_{sky}}{2} \lb2l+1\rb\lb C'^{\bar v' \bar v'}_{{\alpha\alpha l}} \rb^2\rb^{\frac{1}{2}}.\label{s2n_tot}
\ee
Note that the total signal-to-noise is simply the Fisher information, which is a basis-independent quantity unaffected by our PCA; we have confirmed that an identical total signal-to-noise is obtained using the original, non-diagonal basis.

\noindent
\begin{figure}[htb]
\includegraphics[scale=0.5]{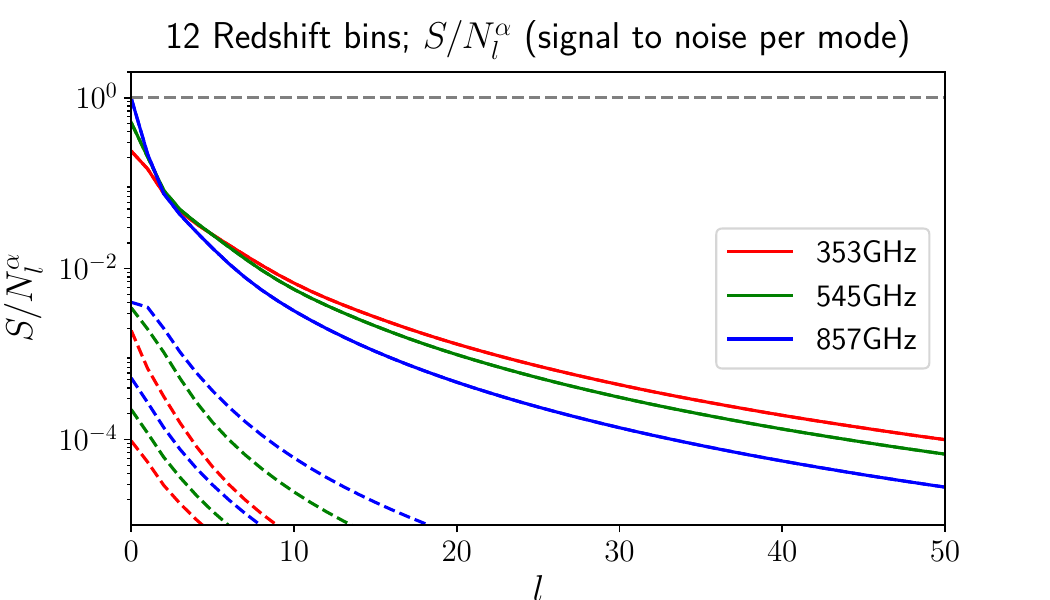}
\caption{The signal-to-noise per mode on the first three principle components (solid, dashed, dot-dashed) of the remote dipole field for Planck-quality data on the full sky. The dashed grey line corresponds to a signal-to-noise of 1. No mode can be reconstructed at signal-to-noise greater than one using Planck.}\label{fig:sn_l_planck}
\end{figure}

\section{Forecast Results}\label{sec:forecast_results}
\subsection{Planck}\label{sec:Planck}

\begin{table*}[htb]
\begin{tabular}{|c|c|c|c|c|}
\hline
Frequency&143  GHz & 353 GHz & 545 GHz &857 GHz\\
\hline
\hline
Noise ($\mu K_{CMB}^2)$&$4.8\times10^{-5}$&$1.30\times10^{-3}$&$4.10\times10^{-2}$&$23.5$\\
\hline
$\Theta_{FWHM}$(arcmin)&7.3&4.94&4.83&4.64\\
\hline%
\end{tabular}
\caption{Noise and resolution for the Planck forecast, taken from~\cite{Aghanim:2018fcm}.}\label{table:noise}
\end{table*}

We first perform a signal-to-noise forecast using values for the experimental noise \eqref{Instrumental_Noise} appropriate for the Planck experiment, to see if there is in principle sufficient statistical power in existing data to reconstruct the remote dipole field using the CIB. As such, we use the specifications for the High-Frequency Instrument from Table 12 of \cite{Aghanim:2018fcm}. We choose the 143-GHz channel for the CMB and we consider the CIB at 353, 545, and 857 GHz. We further assume data on the full sky, neglect foregrounds, and assume that the CIB and CMB can be perfectly separated. Our noise values are given in Table \ref{table:noise}. 

For our redshift-binning scheme, we employ 12 bins of equal comoving width between $0.1< z < 6$. This choice of redshift range includes most of the signal in the CIB and kSZ anisotropies. Increasing the number of bins does not increase the total signal-to-noise. For Planck-quality data, we find that the signal-to-noise per mode is below one for all principle components; see Figure \ref{fig:sn_l_planck}. Summing over all modes and principle components, the total signal-to-noise for the 353, 545, and 857 GHz channels the total signal-to-noise ratios are (0.63, 0.97, 1.5) respectively. We note that the contribution from the monopole ($l=0$) of the reconstructed dipole field accounts for up to half the cumulative signal-to-noise. Since the different frequency bands of the CIB are highly correlated, it is unlikely that these can be combined to reach a signal-to-noise greater than 1. 

\begin{figure}[h]
\includegraphics[scale=0.5]{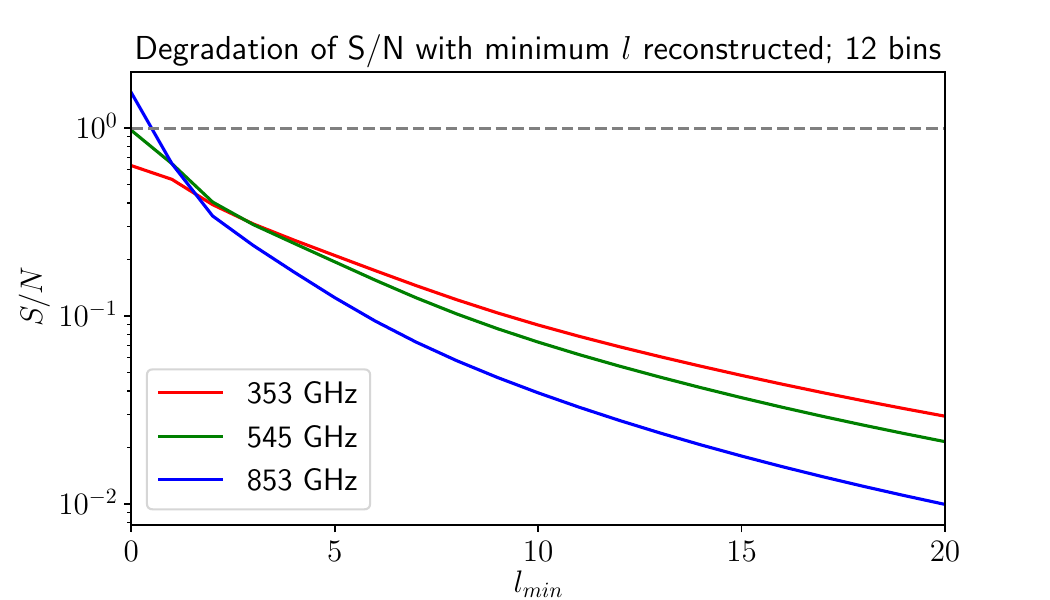}
\caption{Total signal-to-noise as a function of the minimum multipole $l_{\rm min}$ of the first principle component of the remote dipole field that can be reconstructed for Planck-quality data. We see that as we lose access to the remote dipole field on large angular scales (say by having partial-sky data) the total signal-to-noise drops significantly.}\label{fig:minell_planck}
\end{figure}

To date, the largest foreground-cleaned CIB maps~\cite{Lenz:2019ugy} have $f_{\rm sky} \sim 0.25$. Naively, from Eq.~\eqref{s2n_tot}, this reduces the total signal-to-noise by a factor of two. An alternative measure of the degradation in total signal-to-noise due to partial sky coverage is to assume that the dipole field can only be reconstructed above a minimum multipole $\ell_{\rm min}$. In Figure \ref{fig:minell_planck}, we plot the total signal-to-noise as a function of a minimum multipole $\ell_{\rm min}$ in the sum over $l$ in \eqref{s2n_tot}. However, because most of the signal-to-noise is on the largest angular scales, the penalty can be significantly larger than expected from $f_{\rm sky}$, especially at high frequencies. We note, however, that because kSZ tomography uses {\em small angular scale} modes of the CIB to reconstruct the remote dipole field on large angular scales, it may be possible to use less aggressive sky cuts and retain a larger fraction of the CIB than has been used in previous analyses.

\begin{figure}[h]
\includegraphics[scale=0.5]{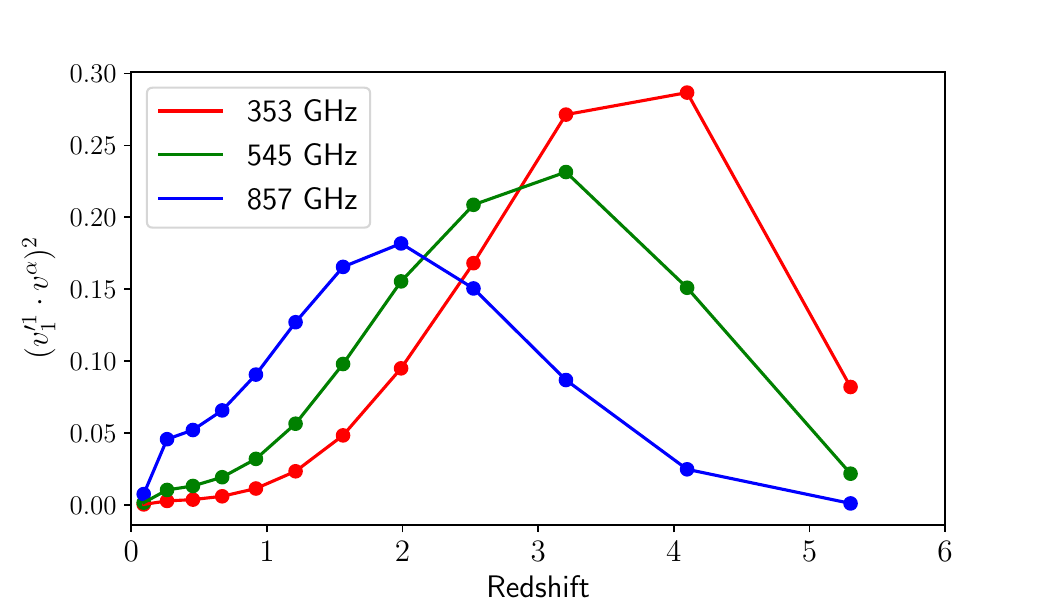}
\includegraphics[scale=0.5]{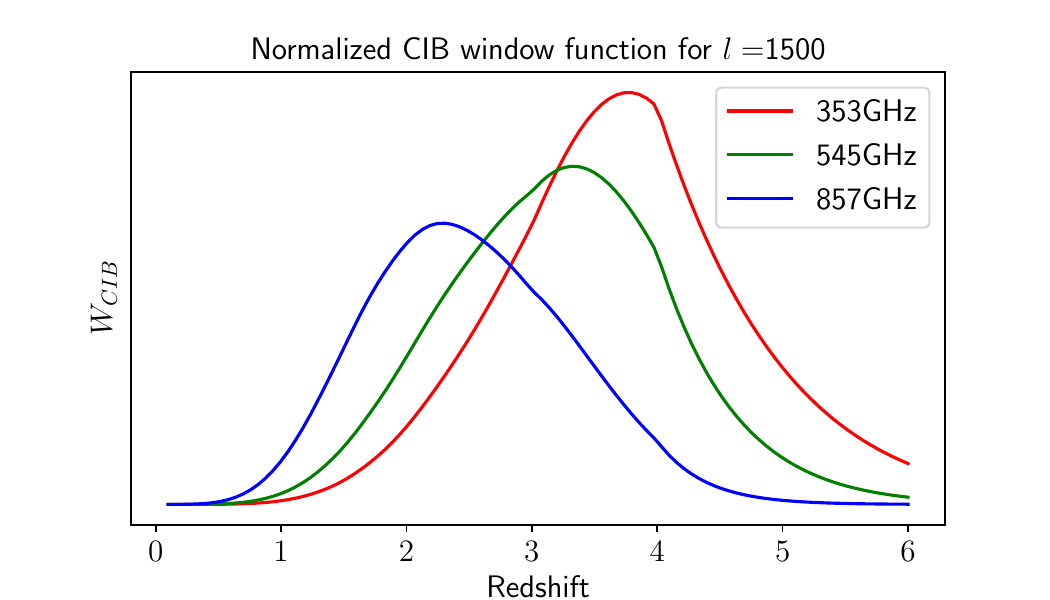}
\caption{The first principle component of the remote dipole field $\hat v'^\alpha$ in the original $\hat v^\alpha$ basis for Planck-quality data at $\ell = 1$ (top panel). For comparison, we plot the power contributed from each redshift to the CIB at $\ell = 1500$ (curves do not strongly depend on the choice of $\ell$), defined by $C_\ell^{II}=\int {d\chi}W_\ell^{CIB}(\chi)$. As expected, the principle components appropriately weight redshift bins according to their contribution to the observed CIB.}\label{fig:components_planck}
\end{figure}

It is interesting that there is one principle component that is reconstructed with significantly higher signal-to-noise than the rest (see Figure \ref{fig:sn_l_planck}). To get insight into what this linear combination is, we can plot the (normalized) components of this mode in the original basis against redshift; see Figure \ref{fig:components_planck}. We also plot the power contributed to the CIB from each redshift $W_\ell^{CIB}(\chi)$, defined by $C_\ell^{II}=\int {d\chi}W_\ell^{CIB}(\chi)$, at the relevant frequencies. The first principle component closely traces $W_\ell^{CIB}(\chi)$, with the lower frequency getting information from more distant redshifts. This is as expected, since the ability to reconstruct the remote dipole field at given redshift using kSZ tomography depends on the presence of measurable CIB fluctuations from that redshift. 

\subsection{Future Experiments}\label{sec:future_experiments}

\noindent
Although Planck-quality data is not sufficient to reconstruct the remote dipole field using the correlation between the CIB and kSZ temperature anisotropies, future experiments stand to greatly improve these measurements. Here, we consider a hypothetical experiment that measures the CMB and CIB at Planck frequencies with noise a factor of 10 lower than the values in Table~\ref{table:noise} at a resolution of 1 arcminute. This is roughly consistent with proposals such as CMB-S4 and CMB-HD. We also lower the flux cuts above which point sources will be removed by a factor of 10 (see Appendix \ref{sec:shot_noise}), thus (in principle) lowering the shot noise on the CIB, although the effect is negligible. For comparison, flux cuts used for SPT analysis~\cite{George:2014oba} are a factor of $\sim 50$ lower than Planck~\cite{2014A&A...571A..30P}.

\begin{figure}[htb]
\includegraphics[scale=0.5]{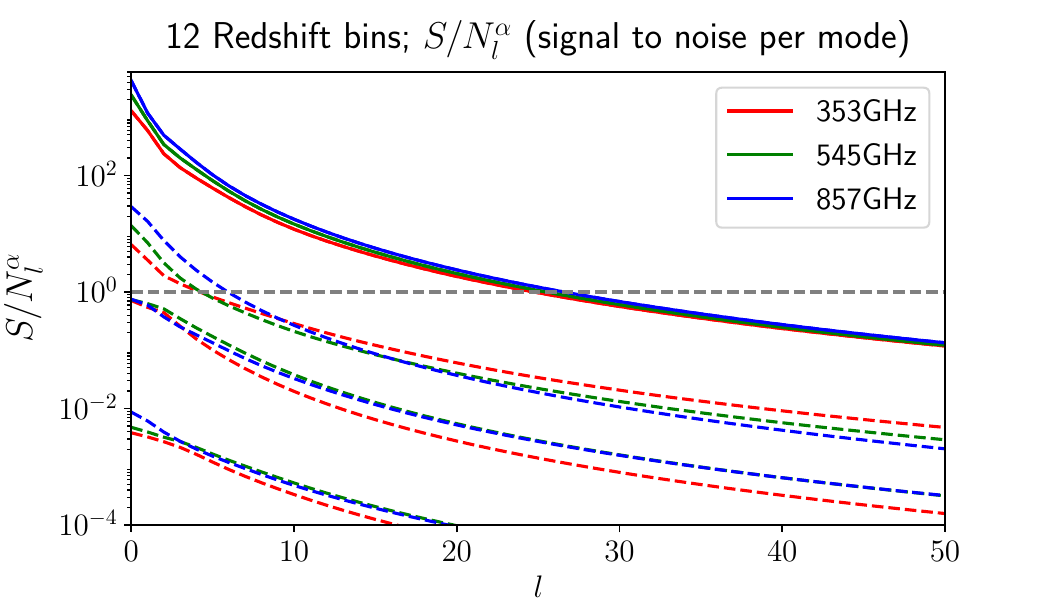}
\caption{Signal-to-noise per mode of the first four principle components of the remote dipole field for a future experiment. For these specifications, it is possible to perform reconstruction at high fidelity for large-angular scale modes of the first two principle components of the remote dipole field. }\label{fig:lower_noise_sn_lalpha}
\end{figure}

\begin{figure}[htb]
\includegraphics[scale=0.5]{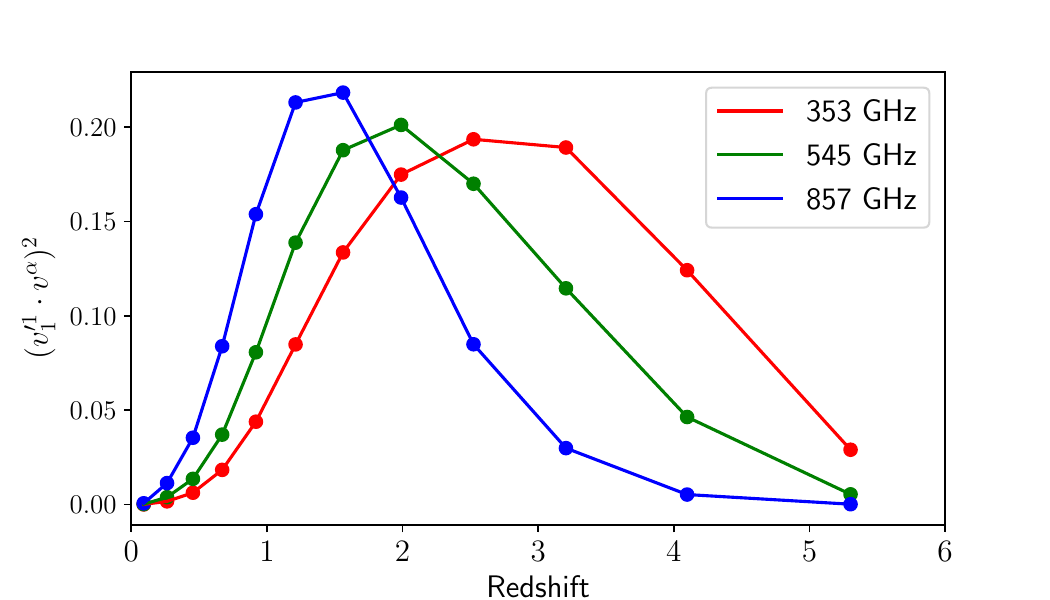}
\caption{ The first principle component of the dipole field computed using noise properties of a future experiment; compare with Figure \ref{fig:components_planck}.}\label{fig:components_future}
\end{figure}

\begin{figure}[htb]
\includegraphics[scale=0.5]{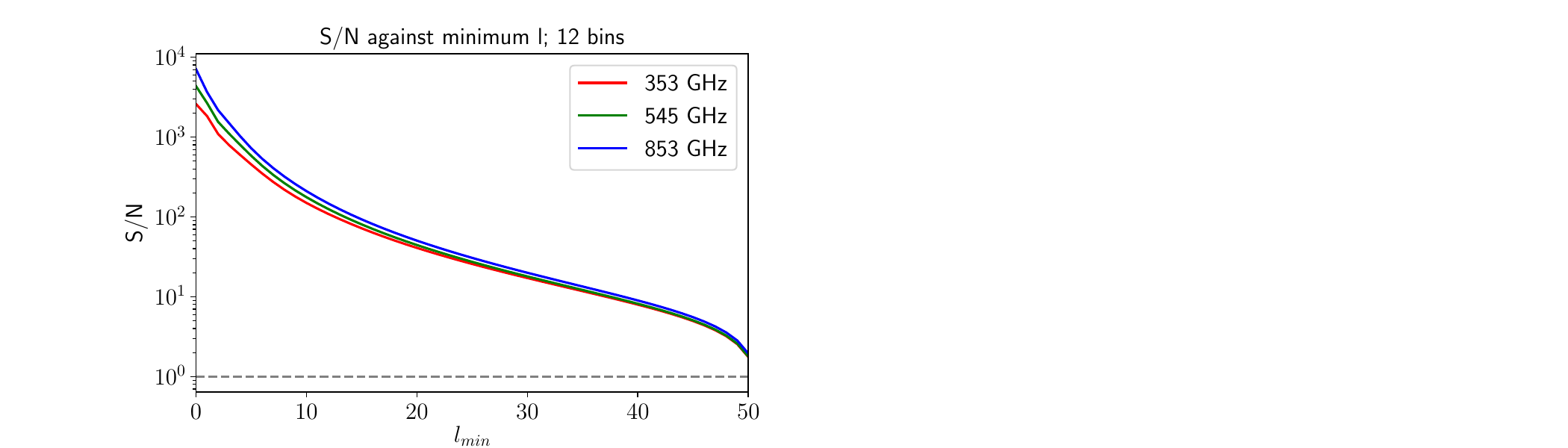}
\caption{Total signal-to-noise for a future experiment as a function of the minimum multipole $\ell_{\rm min}$ that can be reconstructed. This indicates it is possible to make a detection even for significant sky cuts.}\label{fig:total_s2n_lessnoise}
%\tcb{We also include (dashed lines) the fall-off as would be expected from simply multiplying by the sky fraction $f_{sky}$.}}
 \end{figure}

Upon considering these noise specifications, we find clear improvements in our signal-to-noise forecasts. For 12 redshift bins, and assuming full-sky data, no foregrounds, and perfect separation of the CMB and CIB, the total signal-to-noise at (353, 545, 857) GHz goes from (0.63, 0.97, 1.5) to (2600, 4366, 7100). For data of this quality, it is possible to achieve signal-to-noise per mode far greater than one on large angular scales. We plot the signal-to-noise per mode $S/N_l^\alpha$ in Figure \ref{fig:lower_noise_sn_lalpha} for the first four principle components. A high-fidelity map of the first principle component could be reconstructed up to $l\sim 20$. The shape of the first principle component in the redshift basis is shown in Fig.~\ref{fig:components_future}. Comparing with Fig.~\ref{fig:components_planck}, we see that there is relatively more weight at lower redshift for the high frequency channels than for Planck-quality data. We also explore the dependence of the total signal-to-noise on the minimum multipole $\ell_{\rm min}$ that can be reconstructed in Fig.~\ref{fig:total_s2n_lessnoise}. Even for relatively large $\ell_{min}$, it is still possible to obtain a total signal-to-noise greater than one with a future experiment. We conclude that achievable future experiments will in principle have the statistical power to perform kSZ tomography using the CIB. 

The modes which can be reconstructed with the highest fidelity are on the largest angular scales. To properly interpret the reconstruction on these scales, it is important to include all of the contributions to the remote dipole field, as described in Sec.~\ref{sec:reconstruction_estimators}. In particular, it is not a good approximation to replace $v_{\rm eff} \simeq \vec{v} \cdot {\bf \hat{n}}$. For the monopole and dipole of the first principle component, the doppler contribution is of the same order as the SW and ISW terms. The SW and ISW contributions reach the percent-level only for $\ell > 5$. Any analysis using the largest angular scales of the CIB-based reconstruction should therefore include all contributions to the remote dipole field.

\section{Correlations with remote dipole reconstruction from a galaxy redshift survey}\label{sec:correlations}

\subsection{Information content}

Previous work~\cite{Terrana:2016xvc,Deutsch:2017ybc,Smith:2018bpn,Cayuso:2018lhv} demonstrated that a high fidelity reconstruction of the remote dipole field over a range of redshifts will be possible using future CMB experiments in concert with large galaxy redshift surveys, such as LSST. The constraining power of these future measurements for a variety of cosmological scenarios was subsequently explored in Refs.~\cite{Zhang:2015uta,Munchmeyer:2018eey,Madhavacheril:2019buy,Cayuso:2019hen,Contreras:2019bxy,Pan:2019dax}. Given that the remote dipole field will already be reconstructed quite well, it is natural to ask if the very coarse-grained reconstruction provided by the CIB will provide any useful new information. A quantitative measure of the information content in a set of correlated observables is given by the Fisher information:
\begin{equation}\label{eq:fisher_information}
F = \sum_{\ell} \frac{2 \ell+1}{2} {\rm Tr} \left[ {\bf C}_\ell ({\bf C}_\ell + {\bf N}_\ell )^{-1} {\bf C}_\ell ({\bf C}_\ell + {\bf N}_\ell )^{-1} \right].
\end{equation}
The $(N_{\rm bin} +1) \times (N_{\rm bin} +1)$ covariance matrix ${\bf C}_\ell$ includes the auto- and cross-correlation between $v_{\ell m}^\alpha$ in $N_{\rm bin}$ redshift bins reconstructed using a galaxy survey and the first principle component ${v'}_{\ell m}^1$ from the reconstruction using the CIB (other principle components have far lower signal-to-noise). We assume that the noise covariance matrix ${\bf N}_\ell$ is diagonal, with the reconstruction noise on ${v'}_{\ell m}^1$ calculated using the specs for the future experiment from above and reconstruction noise on the $v_{\ell m}^\alpha$ computed using LSST~\cite{2009arXiv0912.0201L} as our proxy for a galaxy survey, as in Ref.~\cite{Deutsch:2017ybc}.~\footnote{The reconstruction noise depends on our model of the galaxy bias and the shot noise for LSST. We assume the galaxy bias is $b =0.95/D(z)$, where $D(z)$ is the growth function, and that the number density of galaxies per arcmin$^2$ is $n(z) = (n_{\rm gal} / 2z_0) (z/z_0)^2 \exp(-z/z_0)$ with $z_0=0.3$ and $n_{\rm gal} = 40 \ {\rm arcmin}^{-2}$ } We project out various observables by sending the corresponding noise to infinity.   

The Fisher information Eq.~\eqref{eq:fisher_information} for the CIB-based reconstruction at 353 GHz is $F_{\rm CIB} = 19.3$. The Fisher information for the galaxy-based reconstruction is significantly larger, $F_{\rm g} = 74.3$, due to the larger number of modes that are reconstructed at appreciable signal-to-noise. If the information in the CIB-based reconstruction was independent of that in the galaxy-based reconstruction (as might be expected from the redshift weighting of the fist principle component, as in Fig.~\ref{fig:components_future}), the combined Fisher information would simply be the sum of these two. However, at the large angular scales on which it can be reconstructed, the remote dipole field has a significant correlation length (see e.g.  Ref.~\cite{Deutsch:2017ybc}). Accounting for these correlations, the Fisher information using the full set of observables is $F_{\rm g+CIB} = 76.5$, implying that only roughly $11\%$ of the information in the CIB-based reconstruction is independent. We therefore expect the CIB-based reconstruction to offer limited improvements in the constraints on cosmological models beyond what is possible using the galaxy-based reconstruction. However, we emphasize that this analysis only accounts for statistical error, and that the systematics associated with the CIB-based reconstruction could be less severe, or complementary, to the galaxy-based reconstruction. In this case, the additional information from the CIB-based reconstruction could be important for deriving cosmological constraints from the remote dipole field. A study of mock data, which we refer to future work, will be able to quantify better the effect of systematics on each case.

\subsection{Optical depth degeneracy}

A significant obstruction to using kSZ tomography for cosmology arises from the inability to perfectly model the correlations between the optical depth and the tracer being used in the reconstruction, in this case the CIB intensity (e.g. the power spectrum Eq.~\eqref{cltauj}, which is a necessary component for the dipole field estimator). This model uncertainty manifests itself as a redshift-dependent linear bias on the reconstructed dipole field, and is known as the ``optical depth degeneracy"; see Refs.~\cite{Battaglia:2016xbi,Smith:2018bpn,Madhavacheril:2019buy} for a detailed discussion. This optical depth bias is degenerate with the amplitude and growth of structure, making it difficult to derive cosmological constraints from the reconstructed dipole field alone. However, we can utilize the fact that both the galaxy-based and CIB-based reconstructions trace the same realization of the remote dipole field to measure the ratio of the optical depth bias of the two tracers as a function of redshift. This is an example of sample variance cancelation~\cite{Seljak:2008xr,McDonald:2008sh}, and in principle the ratio of bias parameters can be measured arbitrarily well in the limit of vanishing reconstruction noise -- e.g. without cosmic variance. 

We can investigate this by considering again the $(N_{\rm bin} +1) \times (N_{\rm bin} +1)$ covariance matrix ${\bf C}_\ell$, with $N_{bins}$ columns corresponding to galaxy reconstruction and the remaining column to the first principle component. We include the optical depth modelling-bias parameters $b_g^\alpha$ and $b_{CIB}^\alpha$ in the covariance matrix using the definitions:
\begin{equation}
v_g^\alpha = b_g^\alpha \bar v ^\alpha, \ \ \ v'_{CIB} = \sum _\alpha c^\alpha b_{CIB}^\alpha \bar v^\alpha
\end{equation}
where $\bar v^\alpha$ is the true dipole field in bin $\alpha$ and $c^\alpha$ is the eigenvector of the first principle component. The introduced optical-depth ``bias'' $b^\alpha$ parameters (which are unrelated to the bias that appears in the power spectrum) quantify modelling uncertainties in the electron---galaxy or electron---CIB cross-corellations.

\begin{figure}[h!]
\includegraphics[scale=0.5]{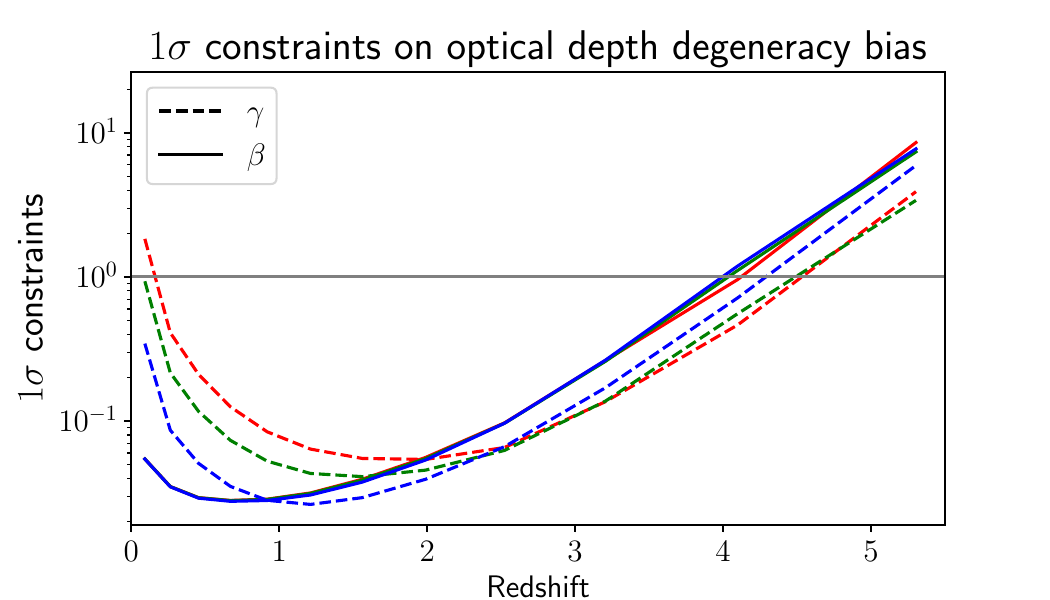}

\caption{{$1 \sigma$ constraints on the optical depth bias.} We plot the results of our Fisher forecast for the $1-\sigma$ constraints on the bias parameters $\beta^\alpha =b_g ^\alpha A_S$ (solid line) and $\gamma^\alpha = b_{CIB}^\alpha / b_g^\alpha$ (dashed line). We performed three separate forecasts, one at each frequency. }\label{fig:bias}
\end{figure}

We consider a simplified analysis, where the amplitude of the primordial power spectrum $A_s$ is allowed to vary, but the other $\Lambda$CDM parameters are held fixed. The bias parameters are each totally degenerate with $A_s$, which is the manifestation of the optical depth degeneracy. We therefore define a reduced parameter space characterized by:
\begin{equation}
\gamma^\alpha \equiv b_{CIB}^\alpha/b_{g}^\alpha \ \ \ \ \beta ^\alpha \equiv b_g^\alpha A_s
\end{equation}
We compute the forecasted 1-sigma constraints on  $\beta^\alpha$ and $\gamma^\alpha$  from the Fisher matrix:
\be
F_{AB}=\sum_{l}\frac{2l+1}{2}\Tr\left[\lb{\bf C}_\ell+{\bf N}_\ell\rb^{-1}\frac{\partial{\bf C}_\ell }{\partial \Pi^A}\lb{\bf C}_\ell+{\bf N}_\ell\rb^{-1}\frac{\partial{\bf C}_\ell }{\partial \Pi^B}\right]
\ee
where $\Pi^A$ is our $2 N_{bin}$-dimensional parameter vector. We use 12 bins and assume fiducial values of $b_g^\alpha =1$, $b_{CIB}^\alpha=1$, and $A_s=2.2$ (the factor of $10^{-9}$ is absorbed into the definition of the dipole field), translating to $\gamma^\alpha = 1$ and $\beta ^\alpha=2.2$. The 1-sigma marginalized constraints $\sqrt{\lb F^{-1}\rb}_{AA}$ for $\beta^\alpha$ and $\gamma^\alpha$ are shown in Fig.~\ref{fig:bias}. We have assumed the noise properties of the future experiment described above, along with galaxy number densities consistent with those expected from LSST. Note that we have not included the covariance between the different frequency channels, presenting each as a separate forecast. For all frequencies, the constraint on $\gamma^\alpha$ is order $\sim 10\%$ in the redshift range $1 \alt z \alt 3$, which is where the galaxy-based reconstruction noise is lowest. The best constraint on $\gamma^\alpha$ reaches the $\sim 3 \%$-level, for the 857 GHz channel at $z \sim 1$. This is expected, since the first principle component of the dipole field reconstructed using the 857 GHz channel has the highest correlation coefficient with the galaxy-based reconstruction, peaking around $z \sim 1$. Percent level constraints on $b_g^\alpha$ can be obtained by correlating the remote dipole field with another tracer, such as the distribution of fast radio bursts~\cite{Madhavacheril:2019buy} or the large-scale modes of a galaxy survey~\cite{Contreras:2019bxy}. It therefore seems likely that percent-level measurements of $b_{CIB}^\alpha$ itself will be possible, allowing for cosmological information to be harvested from the CIB-based dipole field reconstruction. Such measurements would also provide information on galaxy formation and evolution, e.g. through constraints on the parameters in the halo model described in Appendix~\ref{App:halo_CIB}. 

%\clearpage
%%%	CONCLUSION		%%%
\section{Conclusions}\label{sec:conclusions}

\noindent
We have constructed a quadratic estimator for the remote dipole field based on the CIB and CMB temperature, generalizing previous work on kSZ tomography to two-dimensional tracers of large scale structure. Existing datasets of the CMB and CIB nearly have the sensitivity, resolution, and sky coverage to make a statistically significant detection of the remote dipole field. Our forecast for datasets with comparable sensitivity and resolution to the Planck satellite indicate that a detection of signal-to-noise of order one could be made in the absence of foregrounds and sky cuts. However, an idealised future experiment with roughly an order of magnitude better sensitivity, a beam of one arcminute, and lower flux-cut for point-source removal could in principle make a detection with total signal-to-noise of $\mathcal{O}(1000)$. 
Next-generation experiments such as Simons Observatory~\cite{Ade:2018sbj}, CCAT-p~\cite{2018SPIE10700E..5XP}, CMB-S4~\cite{Abazajian:2016yjj}, PICO~\cite{Hanany:2019lle}, or CMB-HD~\cite{Sehgal:2019nmk} fall somewhere between Planck and such a future instrument, making it likely that even with complications such as foreground removal, partial sky coverage, instrumental systematics etc. that a high-fidelity CIB-based reconstruction of the coarse-grained remote dipole field will be achievable. 

Because of the wide redshift window sampled by the CIB at a fixed frequency, it is only possible to reconstruct the remote dipole field averaged over a very large volume. When considering correlations over such large scales, it is not sufficient to approximate the remote dipole field by the local Doppler shift induced by peculiar velocities -- the Sachs Wolfe, Integrated Sachs Wolfe, and primordial Doppler components must be retained. The remote dipole field on such large scales contains information about early-Universe physics, and future experiments could meaningfully constrain a number of scenarios, as considered in Ref.~\cite{Cayuso:2019hen}. The CIB samples a different range of redshifts at different frequencies, allowing the remote dipole field to be reconstructed over different, overlapping volumes/redshifts. Although the reconstructed fields will be significantly correlated, if the CIB could be sampled densely in frequency,  it may be possible to extract some information about the growth rate of structure from the remote dipole field or contribute meaningful constraints on primordial non-Gaussianity~\cite{Munchmeyer:2018eey} and modified gravity~\cite{Pan:2019dax}. In addition, it may be possible to use the reconstructed remote dipole field to isolate General Relativistic corrections to the observed CIB on the largest angular scales~\cite{Tucci:2016hng,Contreras:2019bxy}. 

There is significant model uncertainty in the reconstructed remote dipole field, arising from our imperfect knowledge of the CIB-optical depth cross-spectrum $C_{\ell \alpha}^{I\tau}$ (which is a function of redshift). This manifests itself as a bias on the amplitude of the reconstructed remote dipole field, known as the optical depth bias (see Ref.~\cite{Smith:2018bpn} for a detailed overview). Correlations with a galaxy-based reconstruction of the remote dipole field can be used to constrain the optical depth bias at the $\sim 10\%$-level over a range of redshifts, yielding information on the CIB and distribution of electrons.  

Given the potential high-significance reconstruction of the remote dipole field using kSZ tomography with the CIB, the present investigation motivates preliminary investigation with existing data to obtain constraints, and analysis of future data to obtain high fidelity reconstructions.  

\section{Acknowledgements}
 \noindent
We thank O. Dore, G. Holder, D. Lenz, and M. Munchmeyer for useful discussions. MCJ is supported by the National Science and Engineering Research Council through a Discovery grant. FMcC acknowledges support from the Vanier Canada Graduate Scholarships program. This research was supported in part
by Perimeter Institute for Theoretical Physics. Research at Perimeter Institute is supported by the Government of
Canada through the Department of Innovation, Science and Economic Development Canada and by the Province
of Ontario through the Ministry of Research, Innovation and Science.

\appendix

\section{The Halo Model for the CIB}\label{App:halo_CIB}
\noindent
In modelling the CIB emissivity, we follow the halo model of \cite{2012MNRAS.421.2832S} and use the emissivity model of \cite{Wu:2016vpb}. For details of our HOD see \cite{Smith:2018bpn}; we also use the subhalo mass function of \cite{2010ApJ...719...88T}:
 \be
 \frac{d n}{d  M_{S}}=\frac{0.3}{M_S}\left(\frac{M_{S}}{M}\right)^{-0.7} \exp \left(-9.9\left(\frac{M_{S}}{M}\right)^{2.5}\right).
 \ee
In this Appendix we will now present a brief overview of the model as based on \cite{2012MNRAS.421.2832S,Wu:2016vpb}. 

The CIB intensity density at frequency $\nu$ is a line-of-sight integral over emissivity density $j_\nu$
\be
I_\nu (\hat{\mathbf{ n}})= \int _0^{\chi_{re}}d\chi\, a(\chi )j_\nu\lb\chi, \hat{\mathbf{ n}}\rb;
\ee
$j_\nu$ is an integral over luminosity density
\be
 j_\nu(z)=\int dL_{(1+z)\nu}\frac{d n}{dL_{(1+z)\nu}}\frac{L_{(1+z)\nu}}{4\pi}%=\int dM\frac{dn}{dM}\frac{L_{(1+z)\nu}}{4\pi}
\ee
where $\frac{d n}{dL_\nu}$ is the luminosity function such that $ dL_\nu\frac{d n}{dL_\nu}$ gives the number density of galaxies with luminosity density between $L_\nu$ and $L_\nu+dL_\nu$. The factor of $1+z$ in the frequency accounts for the redshift of the emitted radiation. 

Note that (assuming a monotonic luminosity density-mass relation) we can also consider the mass function $\frac{dn}{dM}$, similarly defined, to go from an integral over luminosity density $dL_\nu$ to an integral over halo mass $dM$.%, and $\frac{dn}{dM}$ is the mass function which is similarly defined.

Within the Limber approximation, the CIB power spectrum is 
\be
C_l^{II,\nu}=\int _0^{\chi_{re}}d\chi\frac{a^2(\chi)}{\chi^2}  \, P_{j,\nu\nu}\lb k=\frac{l+1/2}{\chi},z\rb
\ee
where $P_{j,\nu\nu}(k,z)$ is the emissivity power spectrum; at frequency $\nu$; assuming that the emissivity traces galaxies we can use an emissivity-weighted version of the galaxy power spectrum $P_{gg}$ (which can be modelled within the Halo Model \cite{Cooray:2002dia}); see \cite{2012MNRAS.421.2832S} for more details.

We use the halo model to model $P_{jj}$ and split the correlations into 2-halo and 1-halo terms:
\be
P_{j}=P_{j}^{1h}+P_{j}^{2h}.
\ee

%Indeed, we can convert between flux density on Earth at frequency $\nu$ and emitted luminosity using the relation
%\be
%S_\nu=\frac{L_{(1+z)\nu}}{4\pi\chi^2(1+z)}.
%\ee
\subsection{2-halo term}
\noindent
The 2-halo term of the galaxy power spectrum is a biased tracer of the underlying linear dark matter power spectrum:
\be
P^{2h}_{gg}(k,z)=b^2(z) P_{lin}(k,z).\label{pgg_plin}
\ee
In considering an emissivity-weighted version of $P_{gg}$, this becomes
\ba{
P_j^{2h}(k,z)=\lb\int dM\frac{dn}{dM} b(M,z)\frac{ L_{(1+z)\nu} }{4\pi} \rb^2P_{lin}(k,z).
}
such that in the Limber approximation
\ba{
C_{l}^{II\,2h}=\int d\chi \frac{a^2(\chi)}{\chi^2}\lb\int dM\frac{dn}{dM} b(M,z)\frac{ L_{(1+z)\nu} }{4\pi} \rb^2&\nn\\
\times P_{lin}\lb k=\frac{l+1/2}{\chi},z\rb&.
} 
Note that $L_{(1+z)\nu}$ includes flux both from central galaxies and satellite galaxies: $L_{(1+z)\nu}=L_{(1+z)\nu}^{cen}+L_{(1+z)\nu}^{sat}$. The satellite flux can be found via an integral over subhalo mass $M_S$
\be
L_{(1+z)\nu}^{sat}(M,z)=\int dM_S \frac{dn}{dM_S}L_{(1+z)\nu}(M_S,z)\label{lsat}
\ee
with $\frac{dn}{dM_S}=\frac{dn}{dM_S}(M,M_S)$ the subhalo mass function such that $\frac{dn}{dM_S}dM_S$ gives the number density of subhalos of mass between $M_S$ and $M_S+dM_S$ in a halo of mass $M$. Note that by writing $L_{(1+z)\nu}(M_S,z)$ on the right hand side of equation \eqref{lsat} we are implicitly assuming that, given a luminosity density-(central) halo mass relation $L_{(1+z)\nu}(M,z)$ obeyed by central galaxies, the satellite galaxies will obey the same relation between luminosity density and subhalo mass. 

\subsection{1-halo term}

\noindent
For the one halo term the distinction between satellite and central galaxy is more subtle as we have two types of correlations within a halo: central-satellite and satellite-satellite. 
As such, the 1-halo power spectrum is
\ba{
P^{1h}_j(k,z)=P^{1h}_j(k,z)^{sat-sat}+P^{1h}_j(k,z)^{cen-sat}.
}
These are given by (c.f. the 1-halo dark matter and galaxy power spectra in \cite{Cooray:2002dia}, which here are weighted by emissivity)
\ba{
P^{1h}_j(k,z)=&2\int dM \frac{dn}{dM}\frac{L_{(1+z)\nu}^{cen}}{4\pi}\frac{L_{(1+z)\nu}^{sat}}{4\pi}u\lb k,z\rb+\nn\\
&+\int dM \frac{dn}{dM}\lb\frac{L_{(1+z)\nu}^{sat}}{4\pi}\rb^2u^2\lb k,z\rb
}
with $u(k,z)$ the Fourier-space density profile of the halo (assumed NFW). In the Limber approximation
%\begin{widetext}
\ba{
C_{l}^{II\,1h}=&\int d\chi \frac{a^2(\chi)}{\chi^2}\int dM \frac{dn}{dM}\frac{L_{(1+z)\nu}^{sat}}{4\pi}u\lb k,z\rb\nn\\
&\times\bigg{(}\frac{2L_{(1+z)\nu}^{cen}}{4\pi}+ \frac{L_{(1+z)\nu}^{sat}}{4\pi} u\lb k,z\rb\bigg{)}\bigg{|}_{k=\frac{l+1/2}{\chi}}.\label{CL_1halo}
}

\subsection{Shot Noise Term}\label{sec:shot_noise}

\noindent
The shot noise is due to the discrete nature of the galaxies sourcing the CIB and is given by an intensity-weighted integral over number counts
\be
C_l^{shot-noise}=\int_0^{S_\nu{}_{cut}} S_\nu^2\frac{dN}{dS_\nu}dS_\nu
\ee
where $\frac{dN}{dS_\nu}dS_\nu$ gives the number of galaxies with flux between $S_\nu$ and $S_\nu+dS_\nu$ and ${S_\nu{}_{cut}}$ is a flux cut-off above which point sources can be removed (for the Planck experiment, we take the flux cuts from Table 1 of \cite{2014A&A...571A..30P}). Given a model for luminosity density, we can convert flux density $S_\nu$ to luminosity $L_{(1+z)\nu}$ via
\be
S_\nu=\frac{L_{(1+z)\nu}}{4\pi\chi^2(1+z)}.\label{S_L}
\ee
With this in mind, we can write an estimate for the shot noise as
\be
C_l^{shot-noise}=\int  d\chi\frac{a^2(\chi)}{\chi^2} \int dM \frac{dn}{dM}\lb \frac{L_{(1+z)\nu}}{4\pi}\rb^2,\label{Shot_Noise}
\ee
where we only integrate up to a cut-off luminosity density, which will be $z-$dependent, defined by \eqref{S_L}.
 \subsection{Emissivity Model}
 \noindent
 As well as prescribing a halo model, we must model the luminosity density $L_\nu$.
As such, we use the model of \cite{Wu:2016vpb} which parametrizes the luminosity density as a modified black body
\be
L_\nu=L_{IR}\Theta_\nu
\ee
where $L_{IR}$ is the total infrared luminosity and the spectral energy density  $\Theta_\nu$ is that of a modified black body
\be
\Theta_\nu\propto\nu^{2.1}B_\nu(T_d)
\ee
with $B_\nu$ the Planck function and $T_d$ the dust temperature of the star forming galaxies. $L_{IR}$ is assigned to the star formation rate $SFR$ as
\be
L_{IR}=\frac{SFR}{K_{IR}+K_{UV}10^{-IRX(M_*)}}
\ee
with the IR excess $IRX$ a function of \ti{stellar mass} $M_*$
\be
IRX(M_*)=0.72\log_{10}\lb\frac{M_*}{10^{10.35}M_\odot}\rb+1.32
\ee
and $K_{IR}=1.49 \times 10^{-10}$,  $K_{UV}=1.71 \times 10^{-10}$.
The $SFR-M_*$ relation is given by \cite{Speagle_2014}
\be
SFR(M_*,t)=\lb0.84-0.026 t\rb\log_{10} M_*-\lb6.51-0.11 t\rb
\ee
where $t$ is the age of the Universe in Gyr.

\begin{figure}[h!]
\includegraphics[scale=0.5]{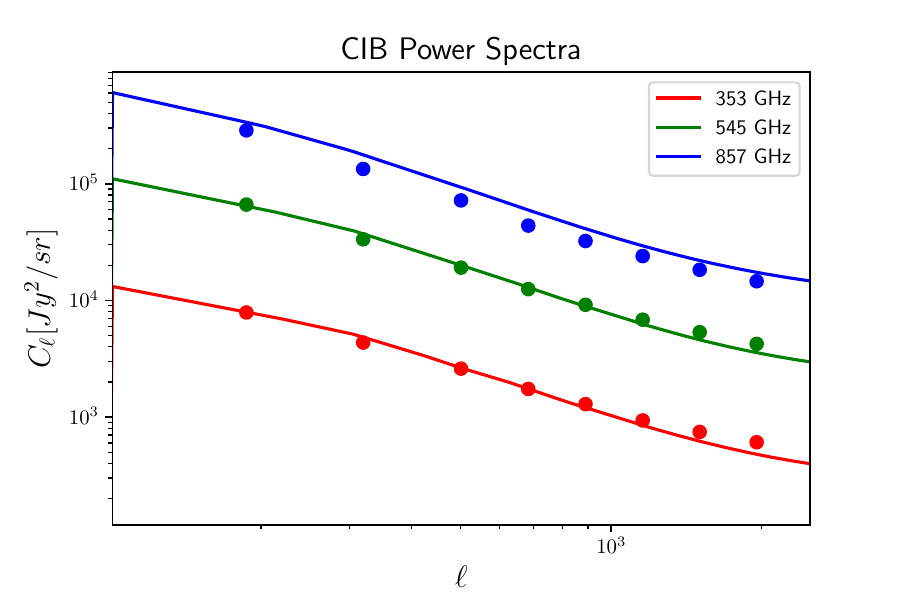}
\caption{The CIB power spectra at 353, 545, 857 GHz. For each frequency, the solid line is our modelled power spectrum, and the points are the data points as measured in \cite{2014A&A...571A..30P}. We note that the data points have been calibrated to a $\nu I_\nu=const$ SED while our computation remains uncalibrated.}\label{fig:CIB_powerspectra}
\end{figure}

As our HOD deals only with \ti{halo} mass $M$, while the luminosity model is dependent on \ti{stellar} mass $M_*$, we must also have a $M-M_*$ relation; for this we perform abundance-matching between our halo mass function $\frac{dn}{dM}$ and the stellar mass functions $\frac{dn}{dM_*}$ as specified in \cite{Wu:2016vpb}:
\ba{
\frac{dn}{d M_*}=\begin{cases}
\lb\Phi_1 m^{1+\alpha_1}+\Phi_2m^{1+\alpha_2}\rb\frac{ e^{-m}}{M_*}\ln 10&0\le z \le 3.5\\
\frac{1}{M_*}\Phi_3m^{1+\alpha_3}\frac{e^{-m}}{M_*}\ln10&3.5<z\le6
\end{cases}
}
with $m\equiv\frac{M_*}{M_0}$ and the parameters given by
\begin{subequations}
\ba{
\log_{10} M_0=&\begin{cases}
10.9+0.08 z&0\le z\le 3.5\\
12.26-0.77\ln\lb1+z\rb&3.5<z\le6;
\end{cases}\\
\log_{10}\Phi_1=&-2.4-0.61 z;\\
\log_{10}\Phi_2=&-3.29-0.23z;\\
\alpha_1=&-0.68;\\
\alpha_2=&-1.57;\\
\log_{10}\Phi_3=&-0.77-1.99\ln\lb1+z\rb;\\
\alpha_3=&-0.47-0.69\ln\lb1+z\rb.
}
\end{subequations}
Armed with this model, we are in a position to plot the anisotropy power spectrum of the CIB; see Figure \ref{fig:CIB_powerspectra}. We compare our model to the Planck data points as given in \cite{2014A&A...571A..30P}. Note however that these data points have been calibrated to a $\nu I_\nu=const$ SED while ours remain uncalibrated.

\subsection{CIB / Electron Cross Power $C_l^{\tau I}$}\label{app:CIB_EL}

\noindent
The total cross-power spectrum between the CIB and the electrons is a sum of a 1-halo and 2-halo term is 
\be
P_{ e j}=P_{e j}(k,z)^{1h}+P_{e j}(k,z)^{2h}
\ee
with  $P_{e j}(k,z)^{1h}$ and $P_{e j}(k,z)^{2h}$ given by
\ba{
P_{e j}(k,z)^{2h}=&\lb\int _0^\infty dM  n(M,z) \lb\frac{M}{\rho_M}\rb b_h(M,z) u_e(k|m,z)\rb\times\nn\\
\times&\lb\int dM\frac{dn}{dM} b(M,z)\frac{ L_{(1+z)\nu} }{4\pi} \rb P_{lin}(k,z)
%&=P_{lin}(k,z)\lb\int _0^\infty dM  n(M,z) \lb\frac{M}{\rho_M}\rb b_h(M,z) u_e(k|m,z)\rb\times \lb\rb.
}
and
\ba{
P_{e j}(k,z)^{1h}=&\int _0 ^\infty dM n(M,z) \frac{M}{\rho_M}u_e(k|M,z) \nn\\
&\times\lb \frac{L_\nu^{cen}(M,z)}{4\pi}+ \frac{L_\nu^{sat}(M,z)}{4\pi}u(k,z)\rb.
}
where $u_e$ is the Fourier space density profile of electrons.

Using the Limber approximation we will get
\be
C_{l\alpha}^{\tau I}=\sigma_T\int _{\chi^\alpha_{min}}^{\chi^\alpha_{max}}d\chi \frac{a^2(\chi)}{\chi^2}P_{e j}\lb k=\frac{l+1/2}{\chi},z\rb.
\ee 
A model must be chosen for the distribution of the electrons $u_e$; we use the `universal' profile of \cite{Komatsu:2001dn}.

\bibliography{references}

\end{document}